# Bound electron nonlinearity beyond the ionization threshold

J. K. Wahlstrand,[1,2] S. Zahedpour,[1] A. Bahl,[3] M. Kolesik,[3] and H. M. Milchberg[1,*]

[1]*Institute for Research in Electronics and Applied Physics, University of Maryland, College Park, MD 20742*
[2]*Engineering Physics Division, National Institute of Standards and Technology, Gaithersburg, MD 20899*
[3]*College of Optical Sciences, University of Arizona, Tucson, AZ 85712*

*Corresponding author: milch@umd.edu

Although high field laser-induced ionization is a fundamental process underlying many applications, there have been no absolute measurements of the nonlinear polarizability of atoms and molecules in the presence of ionization. Such information is crucial, for example, for understanding the propagation of high intensity ultrashort pulses in matter. Here, we present absolute space- and time-resolved measurements of the ultrafast laser-driven nonlinear polarizability in argon, krypton, xenon, nitrogen, and oxygen up to an ionization fraction of a few percent. These measurements enable determination of the non-perturbative bound electron nonlinearity well beyond the ionization threshold, where it is found to be approximately linear in intensity.

Nonlinear polarization driven by strong fields beyond the perturbative limit plays a central role in any experiment or application involving propagation of intense ultrashort optical pulses in material media. It is integral to the process of high harmonic generation and its phase matching [1,2]. It leads to extremely wide bandwidths in supercontinuum generation [3], which can be applied to single cycle pulse generation [4] and ultrafast spectral interferometry [5]. The space- and time-dependence of the nonlinear polarization can also control the collapse and collapse arrest of optical beams in femtosecond filamentation [6,7]. The full nonlinear response of atoms or molecules to intense laser fields is comprised of bound and free electron contributions.

While extensive absolute measurements of the bound nonlinear response of neutral atoms and molecules have been made in noble gases and atmospheric constituents [8-10], there have not been absolute measurements of the free electron contribution to the nonlinear polarization, which requires knowledge of the absolute transient ionization rate. It is such measurements that will complete the picture of the nonlinear polarization in intense non-perturbative fields.

In this Letter, we fully map the complete nonlinear response of several atomic and molecular species through the ionization transition with sufficient accuracy to reveal the bound contribution surviving *above* the ionization threshold. This is made possible by absolute measurements of ultrashort pulse laser-induced transient ionization. We find that in all gases studied, the nearly linear relationship between the bound electronic response and intensity observed at lower intensities [8, 9] extends to intensities where up to a few percent of atoms or molecules are ionized, a region significantly beyond the limits of perturbation theory [11].

Prior measurements of the ionization of atoms and molecules by intense ultrashort optical pulses were performed in vacuum chambers at very low pressure. In these experiments, ionization byproducts (electrons and ions) are directly captured long after the ionizing pulse has passed through the interaction volume, approximately the beam waist region [12, 13], and ionization yields are compared to space and time integrations of ionization rate models. Measurements in atomic hydrogen can be benchmarked to an essentially exact theory [13], while measurements in multi-electron atoms and molecules [12] whose ionization yields are generally plotted in arbitrary units, are scaled to overlie theoretical curves of intensity-dependent rates. However, these measurements do not directly address the absolute nonlinear polarizability of atoms and molecules in the presence of ionization.

In many applications of femtosecond laser pulses, the onset of ionization occurs in the intensity regime where multiphoton ionization transitions to tunneling ionization, as the optical field goes a small perturbation on the atomic potential to being of comparable strength. Because atoms and molecules in this intensity regime are exposed to highly nonperturbative fields, there has been considerable discussion about the effect on nonlinear propagation. In particular, debate has arisen (see, for example, [14] and references therein) regarding potentially exotic contributions [14, 15] to the atomic dipole moment, wherein strongly driven bound electrons have been speculated to exhibit a negative polarizability, with this scenario advanced to explain collapse arrest in femtosecond filamentation [16-18]. This debate has persisted due to both experimental and theoretical complications. Experimentally, the extreme nonlinearity of the response in the presence of ionization means propagation effects on measurements are very difficult to avoid. Theoretically, it has been challenging to cleanly separate the contributions of bound and free electrons [15,19,20]. Here, we present new experimental results showing beyond any doubt that the nonlinear bound response continues to increase approximately linearly with intensity, even in the presence of substantial and increasing ionization.

A diagram of the experiment is shown in Fig. 1. The experiment employs single-shot supercontinuum spectral interferometry (SSSI) [5], which uses broadband supercontinuum (SC) probe and reference pulses to measure the spatio-temporal phase shift in a single shot. The technique provides sub-10 fs time resolution set by the inverse probe bandwidth, and few micron scale spatial resolution [5, 21]. Broadband SC pulses ($\Delta\lambda > 100$ nm) are generated by filamentation in a Xe gas cell and split by a Michelson interferometer into co-propagating probe and reference pulses. The probe, reference, and pump pulses are co-propagated into a thin gas target, with the pump and probe temporally overlapped and the reference preceding them. The effective gas target thickness is $L_{eff} = \int (N(z)/N_0)dz \sim 450\mu m$, where $N(z)$ is the gas density profile along the pump/probe path through the gas tube hole and $N_0$ is the profile mean density [8, 22]. The transient refractive index shift induced by the pump on the medium is encoded on the probe, which is interfered with the reference pulse to form a spectral interferogram in an imaging spectrometer. Analysis of the interferogram yields the phase and amplitude shifts resolved in time and one spatial dimension across the pump spot [5, 21].

In a new scheme, which we call 2D+1 SSSI (the original method [5] can be called 1D+1 SSSI), the pumped spot in the thin gas target is imaged, as in 1D+1 SSSI, to the entrance slit of an imaging spectrometer. We define the axis parallel (perpendicular) to the slit as $x$ ($y$). A motorized scanning mirror downstream of the second imaging lens translates the probe/reference beam image in the $y$ direction.

Each shot captured is a probe/reference interferogram in $\omega$ and $x$ at a particular value of $y$ (= $y_i$, say). Multiple interferograms are averaged before extraction to significantly improve the signal-to-noise ratio [20]. The extracted spectral phase shift $\Delta\phi(x,y_i,\omega)$ is then used to determine $\Delta\Phi(x,y_i,t)$ [21]. A 2D+1 map of the nonlinear response $\Delta\Phi(x,y,t)$ is built by scanning $y$. A total of ~$10^4$ shots are used to construct $\Delta\Phi(x,y,t)$. An important feature of 2D+1 SSSI is that a single 2D phase profile at a particular time slice encodes the nonlinear response over a wide range of intensity, greatly improving the statistics of our measurements.

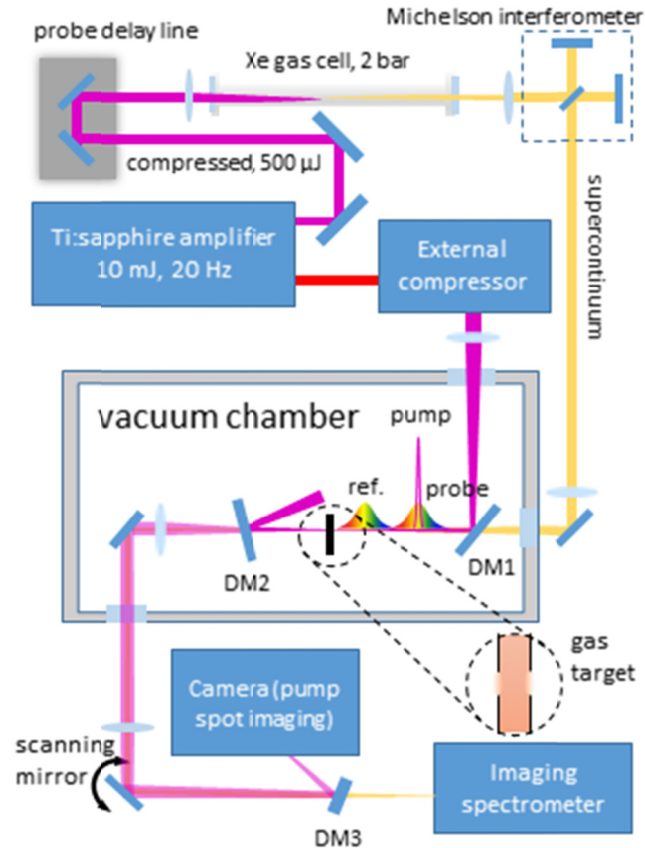

**Figure 1**. Experimental apparatus for 2D+1 SSSI measurement of field-induced ionization. DM1, DM2, DM3: dichroic mirrors. Not shown: 800 nm rejection mirror after Xe gas cell, auxiliary interferometer for gas target characterization, and pre-target pump spot imaging camera.

Figures 2a and 2c show 1D+1 SSSI traces for ionization of $N_2$ and Ar. The more complex $N_2$ trace shows the prompt electronic (Kerr) response, a delayed positive alignment rotational response (which is larger than the Kerr contribution [8]), followed by molecular anti-alignment/alignment and plasma contributions. The Ar trace shows the early Kerr response followed by the rapid tunnel ionization to long-lived plasma. Fig. 2b is a lineout of the Kerr response in Ar below the onset of ionization, which gives the pump intensity envelope. Temporal slices of the extracted 2D+1 phase shift $\Delta\Phi(x,y,t)$ in Ar are shown in Fig. 2e-g for a 42 fs pump pulse of peak intensity 95 TW/cm$^2$ and show the whole beam spatial effect of the positive and negative transient index contributions. Unlike in our previous results [23-25], these new measurements are absolute and quantitative at intensities where ionization is observed. This required increasing the pump spot size to minimize lensing of the pump and probe by the plasma transverse gradient, increasing the time between pump pulses to 100 ms to avoid cumulative thermal effects in the gas [26], and improving the temporal resolution of SSSI by minimizing the probe chirp (consistent with the desired temporal window) and optimizing the spectrometer resolution [21]. Results for the other gases studied, including movies of $\Delta\Phi(x,y,t)$, are provided in [22].

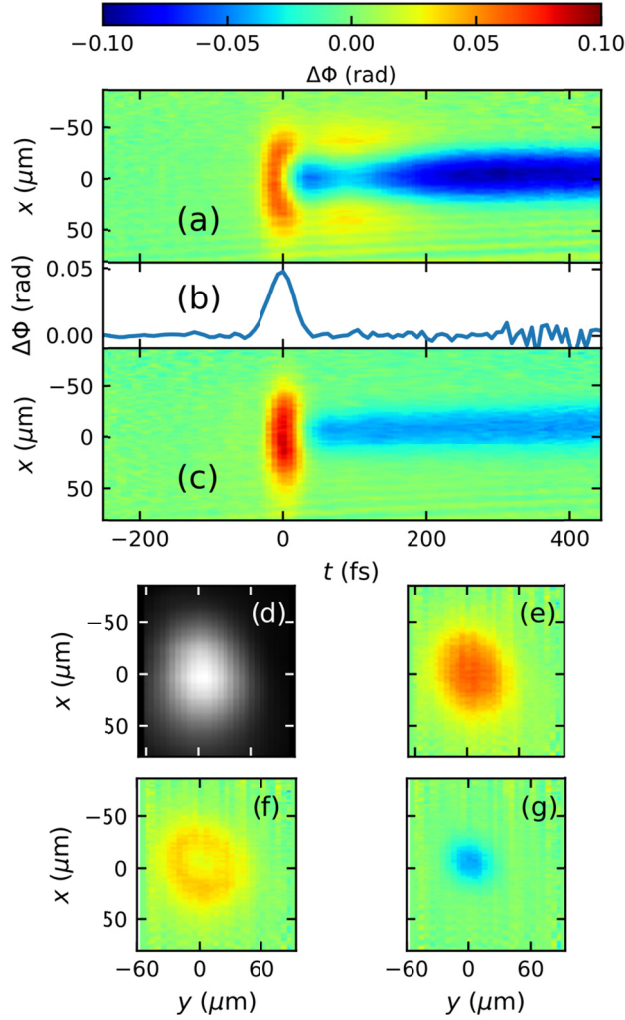

**Figure 2**. Results in Ar for peak pump intensity of 95 TW/cm$^2$ and in N$_2$ for peak pump intensity of 115 TW/cm$^2$. The pulsewidth is 42 fs. (a) 1D+1 phase shift $\Delta\Phi(x,y_0=0,t)$ in N$_2$. The complex structure results from the electronic, rotational, and free electron contributions to the transient polarizability; (b) pump pulse envelope given by $\Delta\Phi(x_0=0,y_0=0,t)$ in Ar at 47 TW/cm$^2$, with the Kerr phase shift proportional to the pump intensity envelope. (c) 1D+1 phase shift $\Delta\Phi(x,y_0=0,t)$ in Ar; (d) Image of the pump spot at the gas target for Ar data. 2D+1 SSSI-derived phase shift images in Ar: (e) $\Delta\Phi(x,y,t=-14$ fs), dominated by the Kerr response, (f) $\Delta\Phi(x,y,t=+25$ fs), showing the Kerr response on the wings and the growing plasma contribution in the center of the beam. (g) $\Delta\Phi(x,y,t=+100$ fs), showing the dominant plasma contribution after the pump pulse. A movie of $\Delta\Phi(x,y,t)$ is provided in [19]. The peak of the pump pulse defines zero for the $x$, $y$, and $t$ coordinates.

We first examine the pure plasma component of the phase shift, which can be isolated by examining time delays long after contributions by the bound electron response, which includes the prompt Kerr response and, in the case of N$_2$ and O$_2$, the delayed rotational response [24]. For our case of a thin gas target in which the probe experiences negligible refraction, the refractive index profile is found using $\Delta n(x,y,t) = \Delta\Phi(x,y,t)/(kL_{eff})$, giving $N_e(x,y) = -2N_{cr}\Delta n(x,y,t_1)$ for the axially averaged electron density profile, where $t_1 > 50$ fs for the noble gases and $t_1 > 250$ fs for $N_2$ and $O_2$. Here we have used the refractive index shift induced by a low density collisionless plasma, $\Delta n = -N_e/2N_{cr}$, where $N_{cr} = 3.1\times10^{21}$ cm$^{-3}$ is the critical electron density at the probe central wavelength $\lambda_{pr} \approx 600$ nm. In the absence of probe refraction, each probe ray centered at $(x_i,y_i)$ samples the dynamics induced by the intensity profile $I(x_i,y_i,t)$. The ionization yield $Y = N_e/N_0$ as a function of intensity is shown in Fig. 3. For each data set (a complete 2D+1 scan of ~$10^4$ consecutive shots at the same nominal peak laser power), the ionization yield data points (for times $t > t_1$) were sorted into 20 intensity bins. The points in Fig. 3 are average values for $Y$ in each bin.

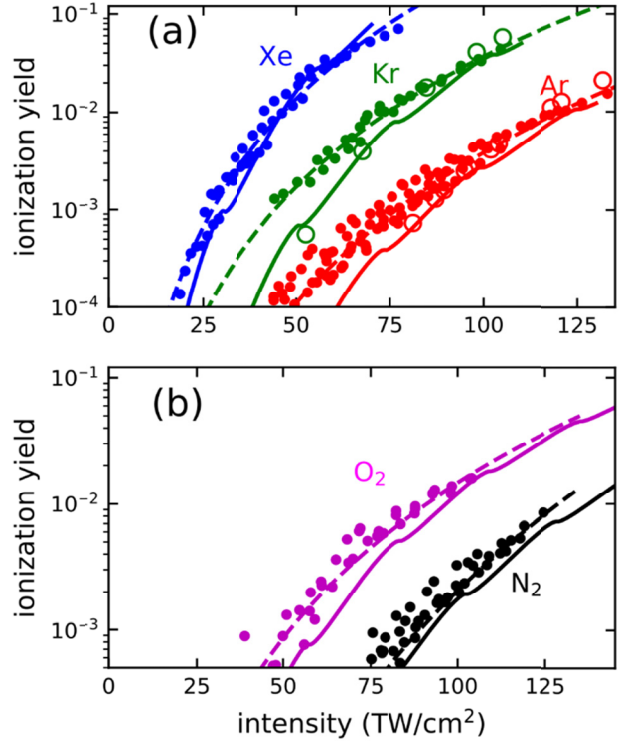

**Figure 3.** Ionization yield (points) as a function of peak intensity $I_0$ for (a) Ar (red), Kr (green), Xe (blue) and (b) $N_2$ (black) and $O_2$ (magenta). PPT [24] rates are shown as solid lines. MESA calculations for Ar and Kr are shown as large open circles [28,29]. Fits to $Y = c_1 I_0^m$, described in [19], are shown as dashed lines, where $c_1$ and $m$ are determined from the fit. The accuracy of the ionization yield measurement is set by the vertical scatter of the points, while the intensity accuracy is ~13% as discussed in the main text.

As in most measurements of the nonlinear response, the error is dominated by the uncertainty in the peak intensity. Here, in a new procedure, we use our previous measurement [8-10] of the nonlinear refractive index of Ar, $n_2 = (9.7\pm1.2)\times10^{-20}$ cm$^2$/W, to provide full 2D calibration of our intensity profiles through $\Delta n(x,y) = 2n_2 I(x,y)$ by directly comparing CCD camera images of the pump spot to 2D Kerr phase shift profiles measured with 2D+1 SSSI at sub-ionization intensities of <50 TW/cm$^2$. The uncertainty in the intensity is 13%, mostly arising from the uncertainty in $n_2$, with the residual uncertainty due to shot-to-shot fluctuations in the measured phase shift of ~3 mrad.

We compare our results to two ionization models. The single-active-electron Peremolov-Popov-Terent'ev (PPT) model [27] (solid lines) shows ionization yields in reasonable agreement with the curves of Fig. 3. We also performed a full simulation of the pump-probe experiment using the unidirectional pulse propagation equation [28, 29] to model pump and probe propagation, and the metastable electronic state approach (MESA) [30-32] to model the full nonlinear response. The results, shown as open circles, are in similarly reasonable agreement with the experimental curves for Ar and Kr [30]. Detailed comparisons of the measured spatiotemporal nonlinear response and MESA simulations, which largely agree, are described in a separate publication [32].

The measured full time-dependent nonlinear response, from the onset of the Kerr response through ionization for argon, is shown in Figs. 4a and 4b. Similar figures for the other gases are found in [22]. At low intensity the response follows the pump pulse intensity envelope, which is well-fit by $I(t) = I_0 e^{-t^2/\tau^2}$ where $\tau = \tau_{FWHM}/(2\sqrt{\ln 2})$ corresponds to our pulse full width at half maximum (FWHM) $\tau_{FWHM}$ = 42 fs (Fig. 4a). Figure 4b shows the time-dependent refractive index shift in Ar for increasing intensity and ionization levels, along with fits to $\Delta n(t) = \Delta n_K e^{-t^2/\tau^2} + \Delta n_p (1+\text{erf}[m^{1/2}t/\tau])/2$, where $\Delta n_K$ is a fitting parameter and $\Delta n_p$ is the peak plasma index shift. The first term is the Kerr response for a Gaussian pulse of peak intensity $I_0$, where $\Delta n_K = 2n_2 I_0$ is the peak index shift experienced by a probe pulse [8]. The second term models the plasma contribution as $N_e(t)/N_0 = Y(t) \approx \int_{-\infty}^{t} w(t')dt'$ using an ionization rate $w(t) = c_2(I_0 e^{-t^2/\tau^2})^m$, for which the yield is $Y(t \to \infty) = c_1 I_0^m$, and where $c_1$ and $m$ are determined from fits to the ionization curves in Fig. 3 and $\Delta n_p = -N_0 c_1 I_0^m (2N_{cr})^{-1}$ [22]. The approximate reduction in Kerr response due to the reduction of the neutral atom density by ionization is accounted for by multiplying the $\Delta n_K$ value found from the fit by 1-$Y$/2. This adjustment, which assumes that the Kerr response from the ions is negligible, reduces $\Delta n_K$ by at most 3% at the highest intensity. This simple model is seen to be an excellent fit to the measured transient index shift. The point of the expression used for $Y$ is not to advance a multiphoton-ionization (MPI)-like model for ionization; it is to provide an analytic model fit to the ionization yield data to enable separation of the bound and free electron contributions. In fact, as seen in [22], the best fit values for $m$ are notably smaller than their corresponding MPI values for each species, indicating the significant contribution of tunneling ionization.

Figures 4c and 4d plot, as a function of peak intensity, the peak Kerr index shift $\Delta n_K$ and the peak plasma shift $\Delta n_p$ extracted from fits to transient index data, as in Fig. 4b, for each species, with Fig. 4c showing atomic results and Fig. 4d showing molecular results. Remarkably, it is seen for all species that $\Delta n_K$ continues to rise with intensity even as $\Delta n_p$ becomes increasingly negative up to the limit of our measurements (at an ionization level of ~5%), *significantly beyond the ionization threshold*. In Ar, Kr, and Xe, the Kerr index shift observed above the ionization threshold is somewhat higher than the curve extrapolated from below the threshold [9], shown as a dashed line in Fig. 4cd. Apparently, $\Delta n_K$ rises slightly faster than linearly in $I$ in the noble gases, but confirmation awaits more accurate measurements. In the molecular gases, the result is somewhat different. In $N_2$, the Kerr index shift above the ionization threshold closely follows the curve extrapolated from the sub-threshold response [8], while in $O_2$ it is below the extrapolated curve. In these gases, the analysis is complicated by the need to separate the bound electronic and rotational responses [22].

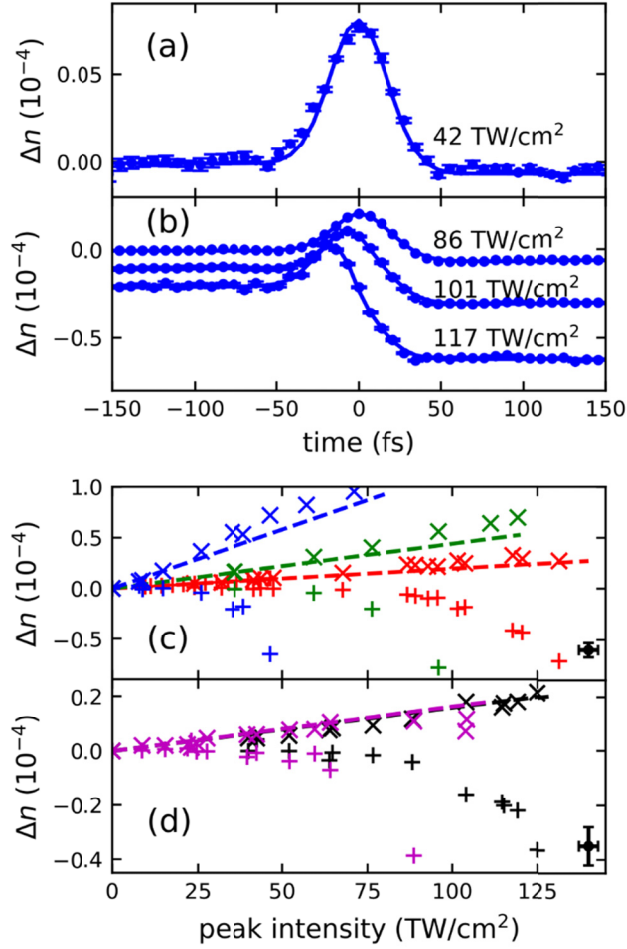

**Figure 4.** Response during pump pulse. (a) Nonlinear index shift in Ar vs. time for peak intensity 42 TW/cm² (dots), below the threshold for ionization, fit to a Gaussian pulse with $\tau_{FWHM}$ =42 fs (solid line). (b) Nonlinear index shift vs. time and fits to the standard model (Kerr effect plus ionization) for Ar (solid lines). The curves have been offset vertically for clarity. (c) Plot of Kerr index change $\Delta n_K$ (×) and plasma index change $\Delta n_p$ (+) from fits as a function of intensity in Ar (red), Kr (green), and Xe (blue). (d) Same as part (c) for $N_2$ (black) and $O_2$ (magenta). The black points on the lower right of panels (c,d) indicate typical uncertainty. Dashed lines show the extrapolated Kerr response $2n_2I$ from previous low intensity measurements [8,9].

That the simple field-quadratic response of bound electrons continues to apply well beyond the ionization threshold is consistent with our Kramers-Kronig (KK) simulations presented in [9]. A physical interpretation of the KK results [9, 33] is that in atoms dressed by the intense field, the change to the single photon absorption coefficient (which contributes to the imaginary part of the effective susceptibility) is non-negligible and dominated by strong ac Stark shifts, which then causes the real nonlinear response (the real part of the effective susceptibility) to be quadratic in the field. This also applies to enhanced 2-photon absorption from resonantly populated high lying states. For non-ground state levels, the range of shifts can be large as the ponderomotive energy, $U_p \sim 8$ eV at 120 TW/cm$^2$, and population can be easily resonantly transferred to states within one or two photons from the continuum [34].

In summary, absolute measurements of ionization in Ar, Kr, Xe, N$_2$, and O$_2$ have enabled absolute determination of the transient free and bound electron contributions to the nonlinear polarizability, applicable to many propagation scenarios involving ionization. For a ~40 fs pump pulse at $\lambda = 800$ nm, the bound component of the nonlinear polarizability is, to within our measurement accuracy, linear in the cycle-averaged intensity over the full range of the interaction up to >100 TW/cm$^2$, which is well past the ionization threshold of the gases measured here, and manifestly in the non-perturbative regime.

The authors thank I. Larkin, E. Rosenthal, N. Jhajj and K.Y. Kim for discussions and technical assistance.

JKW, SZ, and HMM acknowledge support by the Air Force Office of Scientific Research (FA95501610284, FA95501610121); the Army Research Office (W911NF1410372), the Office of Naval Research (N00014-17-1-2705), and the National Science Foundation (PHY1301948). AB and MK acknowledge support by the Air Force Office of Scientific Research (FA95501610121).

**REFERENCES**

1. C. G. Durfee III, A. R. Rundquist, S. Backus, C. Herne, M. M. Murnane, and H. C. Kapteyn, Phys. Rev. Lett. **83**, 2187 (1999).
2. T. Popmintchev, M.-C. Chen, D. Popmintchev, P. Arpin, S. Brown, S. Alisauskas, G. Andriukaitis, T. Balciunas, O. D. Mucke, A. Pugslys, A. Baltuska, B. Shim, S. E. Schrauth, A. Gaeta, C. Hernandez-Garcia, L. Plaja, A. Becker, A. Jaron-Becker, M. M. Murnane, H. C. Kapteyn, Science **336**, 1287-1291 (2012).
3. P. B. Corkum, C. Rolland, and T. Srinivasan-Rao, Phys. Rev. Lett. **57**, 2268 (1986).
4. A. Couairon, M. Franco, A. Mysyrowicz, J. Biegert, and U. Keller, Opt. Lett. **30**, 2657 (2005).
5. K. Y. Kim, I. Alexeev, and H. M. Milchberg, Appl. Phys. Lett. **81**, 4124 (2002).
6. A. Couairon and A. Mysyrowicz, Phys. Rep. **441**, 47 (2007).
7. L. Berge, S. Skupin, R. Nuter, J. Kasparian, and J.-P. Wolf, Rep. Prog. Phys. **70**, 1633 (2007).
8. J. K. Wahlstrand, Y.-H. Cheng, and H. M. Milchberg, Phys. Rev. A **85**, 043820 (2012).
9. J. K. Wahlstrand, Y.-H. Cheng, and H. M. Milchberg, Phys. Rev. Lett. **109**, 113904 (2012).
10. S. Zahedpour, J. K. Wahlstrand, and H. M. Milchberg, Opt. Lett. **40**, 5794-5797 (2015).
11. A. Spott, A. Jaron-Becker, and A. Becker, Phys. Rev. A **90**, 013426 (2014).
12. B. Walker, B. Sheehy, L. F. DiMauro, P. Agostini, K. J. Schafer, and K. C. Kulander, Phys. Rev. Lett. **73**, 1227 (1994).

13. W. C. Wallace, O. Ghafur, C. Khurmi, S. Sainadh U, J. E. Calvert, D. E. Laban, M. G. Pullen, K. Bartschat, A. N. Grum-Grzhimailo, D. Wells, H. M. Quiney, X. M. Tong, I. V. Litvinyuk, R. T. Sang, and D. Kielpinski, Phys. Rev. Lett. **117**, 053001 (2016).
14. M. Richter, S. Patchkovskii, F. Morales, O. Smirnova, and M. Ivanov, New J. Phys. **15**, 083012 (2013).
15. P. Bejot, E. Cormier, E. Hertz, B. Lavorel, J. Kasparian, J.-P. Wolf, and O. Faucher, Phys. Rev. Lett. **110**, 043902 (2013).
16. V. Loriot, E. Hertz, O. Faucher, and B. Lavorel, Opt. Express **17**, 13429-13434 (2009).
17. P. Bejot, J. Kasparian, S. Henin, V. Loriot, T. Vieillard, E. Hertz, O. Faucher, B. Lavorel, and J.-P. Wolf, Phys. Rev. Lett. **104**, 103903 (2010).
18. C. Bree, A. Demircan, and G. Steinmeyer, Phys. Rev. Lett. **106**, 183902 (2011).
19. M. Nurhuda, A. Suda, and K. Midorikawa, New J. Phys. **10**, 053006 (2008).
20. C. Kohler, R. Guichard, E. Lorin, S. Chelkowski, A. D. Bandrauk, L. Berge, and S. Skupin, Phys. Rev. A **87**, 043811 (2013).
21. J. K. Wahlstrand, S. Zahedpour, and H. M. Milchberg, J. Opt. Soc. Am. B **33**, 1476-1481 (2016).
22. See Supplemental Material at [APS - Physical Review Letters link ( )] or at [http://lasermatter.umd.edu/publications.html#bound_electron_supplementary ] for a discussion of experimental details, additional results for Kr, Xe, $N_2$, and $O_2$, and movies of 2D+1 SSSI.
23. Y.-H. Chen, S. Varma, I. Alexeev, and H. M. Milchberg, Opt. Express **15**, 7458 (2007).
24. Y.-H. Chen, S. Varma, A. York, and H. M. Milchberg, Opt. Express **15**, 11341 (2007).
25. J. K. Wahlstrand, Y.-H. Cheng, Y.-H. Chen, and H. M. Milchberg, Phys. Rev. Lett. **107**, 103901 (2011).
26. Y.-H. Cheng, N. Jhajj, J. K. Wahlstrand, and H. M. Milchberg, Opt. Express **21**, 4740 (2013).
27. A.M. Perelomov, V.S. Popov, M.V. Terent'ev, Sov. Phys. JETP **23**, 924 (1966).
28. M. Kolesik and J. V. Moloney, Phys. Rev. E **70**, 036604 (2004).
29. J. Andreasen and M. Kolesik, Phys. Rev. E **86**, 036706 (2012).
30. A. Bahl, E. M. Wright, and M. Kolesik, Phys. Rev. A **94**, 023850 (2016).
31. M. Kolesik, J. M. Brown, A. Teleki, P. Jakobsen, J. V. Moloney, and E. M. Wright, Optica **1**, 323 (2014).
32. A. Bahl, J. K. Wahlstrand, S. Zahedpour, H. M. Milchberg, and M. Kolesik, Phys. Rev. A **96**, 043867 (2017).
33. M. Sheik-Bahae, D. C. Hutchings, D. J. Hagan, and E.W. Van Stryland, IEEE J. Quantum Electron. **27**, 1296 (1991)
34. M. P. de Boer and H. G. Muller, Phys. Rev. Lett. **68**, 2747 (1992).